\documentclass[authoryear,12pt,3p]{jowarticle}
\usepackage{graphicx}
\usepackage{amsthm,amsmath}
\usepackage{amssymb}
\usepackage{amsfonts}
\usepackage{natbib}
\usepackage{setspace}
\usepackage[all]{xy}
\usepackage{enumitem}
\usepackage{titlesec}
\usepackage{mathrsfs}
\makeatletter
\renewcommand\section{\@startsection{section}{1}{\z@}{-3.25ex plus -1ex minus -.2ex}{1.5ex plus .2ex}{\normalsize\bf}}
\renewcommand\subsection{\@startsection{subsection}{2}{\z@}{-3.25ex plus -1ex minus -.2ex}{1.5ex plus .2ex}{\normalsize\bf}}
\renewcommand\subsubsection{\@startsection{subsubsection}{3}{\z@}{-3.25ex plus -1ex minus -.2ex}{1.5ex plus .2ex}{\normalsize\bf}}
\makeatother

\newtheorem{innercustomgeneric}{\customgenericname}
\providecommand{\customgenericname}{}
\newcommand{\newcustomtheorem}[2]{%
  \newenvironment{#1}[1]
  {%
   \renewcommand\customgenericname{#2}%
   \renewcommand\theinnercustomgeneric{##1}%
   \innercustomgeneric
  }
  {\endinnercustomgeneric}
}

\newtheorem{thm}{Theorem}

\newtheorem{prop}[thm]{Proposition}

\newcustomtheorem{prop*}{Proposition}
\newcustomtheorem{cor*}{Corollary}
\newcommand{\norm}[1]{\left\Vert#1\right\Vert}

\newcommand{\Reals}{\mathbb {R}}

\begin{document}
\begin{frontmatter}
\title{Why Be Regular?, Part I}

\author{Benjamin Feintzeig}\ead{bfeintze@uw.edu}
\address{Department of Philosophy \\ University of Washington}
\author{JB (Le)Manchak}\ead{jmanchak@uci.edu}
\author{Sarita Rosenstock}\ead{rosensts@uci.edu}
\author{James Owen Weatherall}\ead{weatherj@uci.edu}
\address{Department of Logic and Philosophy of Science\\ University of California, Irvine}
\begin{abstract}We provide a novel perspective on ``regularity'' as a property of representations of the Weyl algebra.  We first critique a proposal by Halvorson [2004, ``Complementarity of representations in quantum mechanics", \textit{Studies in History and Philosophy of Modern Physics} \textbf{35}(1), pp. 45--56], who argues that the \emph{non-regular} ``position'' and ``momentum'' representations of the Weyl algebra demonstrate that a quantum mechanical particle can have definite values for position or momentum, contrary to a widespread view.  We show that there are obstacles to such an intepretation of non-regular representations.  In Part II, we propose a justification for focusing on regular representations, \emph{pace} Halvorson, by drawing on algebraic methods.\end{abstract}
\end{frontmatter}
\doublespacing
\section{Introduction}\label{introduction}

It is standard dogma that, according to quantum mechanics, a particle does not, and indeed cannot, have a precise value for its position or for its momentum.  The reason is that in the standard Hilbert space representation for a free particle---the so-called \emph{Schr\"odinger Representation} of the Weyl form of the canonical commutation relations (CCRs)---there are no eigenstates for the position and momentum magnitudes, $P$ and $Q$; the claim follows immediately, from this and standard interpretational principles.\footnote{Namely, the \emph{Eigenstate--Eigenvalue link}, according to which a system has an exact value of a given property if and only if its state is an eigenstate of the operator associated with that property.  The Eigenstate-Eigenvalue link has a long and distinguished history \citep{Gilton}, but it is not without its detractors \citep{Fine,Wallace}.}  

But not everyone accepts this simple argument.  In particular, Hans \citet{HalvorsonBohr} has argued that there is, after all, a sense in which a quantum particle may have a definite position or momentum.
As Halvorson points out, the Schr\"odinger representation is not the only possible representation of the Weyl CCRs; there are other representations available, and in these other representations one can have eigenstates for a position operator or a momentum operator.  These alternative representations have been largely neglected because there is a theorem, known as the Stone-von Neumann theorem, which implies that there is a unique representation (up to unitary equivalence) of the Weyl CCRs  with the property of \emph{regularity}.  Since regularity is often taken to be a desirable or important feature of a representation, many physicists and philosophers of physics take the Stone-von Neumann theorem to establish uniqueness, full stop.  But, Halvorson contends, requiring regularity is precisely to beg the question against the advocate of definite momentum and position states.  And insofar as the latter are of conceptual significance, regularity is a suspect assumption.

Halvorson goes on to argue that these alternative position and momentum representations \emph{are} significant: in particular, they make precise certain insights often attributed to Niels Bohr, related to the so-called \emph{Principle of Complementarity}.  According to this principle, a particle may be said to have a definite, precise position; or it may be said to have a definite momentum.  But it cannot have both.  And this is precisely what one finds with the position and momentum representations: in the position representation there exists a position operator with eigenstates, but there does not exist a momentum operator at all.  And vice versa for the momentum representation.  On this reconstruction, then, Bohr's principle identifies a formal feature of our descriptions of quantum systems, related to a limitation on our ability to represent, once and for all, all of the properties we take a (classical) particle to have.

We will not engage further with Halvorson's arguments concerning Bohr.\footnote{Halvorson himself does not engage in Bohr exegesis in his paper.  Bohr is not even cited!}  Instead, we take Halvorson to have raised an important question concerning the status and interpretation of the regularity assumption necessary for the Stone-von Neumann theorem.\footnote{Actually, if one relaxes the regularity condition only slightly, then one can still arrive at a generalization of the Stone-von Neumann theorem; see \citet{CaMoSt99}.  However, now instead of having a \emph{unique} representation, one only arrives at a classification of the inequivalent representations, including the non-regular ones.}  Our goal is to address this question.  We do so in two papers, of which this one is the first.\footnote{The second is  \citep{Feintzeig+WeatherallPart2}.  Both papers are written to be read independently, though they complement one another.  Aside from length and partial independence of the arguments, one reason for splitting the paper into two parts is that the present paper requires less (and different) technicalia than the sequel, and so the two articles present matters at somewhat different levels.}   

In the present paper, we offer a critical response to Halvorson's proposal.  We first argue that admitting non-regular Hilbert space representations of the Weyl CCRs as somehow on par with the standard Schr\"odinger representation leads to difficulties, particularly once one considers dynamics.  We then argue that there is a more charitable interpretation of Halvorson's view on which one steps back from the representations and considers the relationship between regular and non-regular states on a particular abstract algebra.  We conclude this paper by arguing that whether one should take Halvorson's definite position and momentum states to have physical significance depends sensitively on which algebra one takes to correctly represent the physical magnitudes associated with a quantum particle.  In the sequel, we provide an argument that one should adopt a different algebra, and then argue that the considerations motivating this alternative choice provide independent grounds for taking the states and quantities of the Schr\"odinger representation to be the physically significant ones.\footnote{While we do not address these ideas in the present manuscript, our discussion in the sequel will also engage with the ideas Halvorson explores in an earlier paper \citep{HalvorsonTeller}, presented in response to \citet{Teller}; and with the response to Halvorson by \citet{Ruetsche}.}

We begin, in section \ref{prelim}, by presenting some mathematical background concerning the Weyl CCRs and the Schr\"odinger representation.   In section \ref{H} we describe the non-regular representations Halvorson discusses.  We then present, in section \ref{MM}, some difficulties that arise if one tries to understand the position and momentum representations as together describing the space of physically possible states, as one might usually interpret a Hilbert space representation of a quantum system.  In section \ref{F}, we present a different reading of Halvorson's proposal that we take to be more promising, and then sketch an argument that the success of this proposal turns on Halvorson's choice of algebra.  We conclude in section \ref{conclusion} with a discussion of what has been accomplished in this paper, and a description of what we believe remains to be done in the sequel.

\section{The Weyl CCRS and the Schr\"odinger Representation}\label{prelim}

The standard procedure for constructing a quantum theory of some physical system---at least one with finitely many degrees to freedom---is to identify a Hilbert space $\mathscr{H}$ of (pure) states of the system, along with some *-algebra of operators on that Hilbert space, the self-adjoint elements of which represent the physical magnitudes associated with the system.  Generally this *-algebra is determined by requiring that it be generated by (or at least, contain) some collection of operators that together satisfy specified algebraic relations, known as \emph{canonical commutation relations} (CCRs), that are expected to hold between the physical magnitudes defining the system in question.  

For instance, it is natural to take the quantities \emph{position} and \emph{momentum} to be the defining magnitudes for a single particle moving in one dimension.  (We limit attention to a particle in one dimension in what follows for simplicity, though one can extend these arguments to more general cases.)  Following Dirac and others, building on an analogy with the Poisson bracket for position and momentum in a classical description, one requires that any operators $Q$ and $P$ representing position and momentum, respectively, must satisfy the following CCR:
\begin{equation}\label{CCR}
[Q,P] = iI,
\end{equation}
where the brackets are the commutator, $I$ is the identity operator, and where we are working in units where $\hbar=1$.  Thus, to represent a single particle in one dimension, one would like to find a Hilbert space $\mathscr{H}$ and $*$-algebra of operators acting on $\mathscr{H}$ containing self-adjoint operators $P$ and $Q$ satisfying this relation.

This can be done \citep{DuHeSm00}---but it turns out to be inconvenient on technical grounds.  The reason is that operators $Q$ and $P$ on any Hilbert space $\mathscr{H}$ satisfying the CCR can be shown to be, in general, \emph{unbounded}, which means that for any number $K\geq 0$, one can always find a unit vector $\varphi$ in $\mathscr{H}$ such that $||Q\varphi||>K$ (and likewise for $P$).  Working with unbounded operators adds many complications, including, for instance, that they are not defined on all vectors in the Hilbert space on which they act, and that the collection of all unbounded operators on a Hilbert space do not generally form a vector space, much less an algebra.  Bounded operators, by contrast, are far more convenient to work with.  

There is, however, a trick---originally due to \citet{Weyl}---that one can perform to get around introducing unbounded operators, at least in the first instance.  Consider, instead of $Q$ and $P$, one parameter groups $a\mapsto U_a$ and $b\mapsto V_b$.  We think of $U_a$ and $V_b$ as ``formal exponentiations'' of $Q$ and $P$ respectively, so that for any $a\in\mathbb{R}$, $U_a = e^{iaQ}$, and for any $b\in \mathbb{R}$, $V_b=e^{ibP}$.  Then, by the formal properties of the exponential we should expect that if $Q$ and $P$ satisfy the CCR above, $U_a$ and $V_b$ should satisfy:
\begin{equation}\label{WeylForm}
U_aV_b = e^{-iab}V_bU_a
\end{equation}
for all $a,b\in \mathbb{R}$.  The relationship expressed in Eq. \eqref{WeylForm} is known as the  \emph{Weyl form of the CCRs} \citep{Pe90,ClHa01}.  Unlike the operators $Q$ and $P$, one would expect operators $U_a$ and $V_b$ satisfying Eq. \eqref{WeylForm} to be bounded, again by formal properties of the exponential, and so it should be possible to find operators in $\mathcal{B}(\mathscr{H})$ that satisfy the Weyl CCRs.

And indeed, it is possible.  Let $\mu$ be the Lebesgue measure on $\mathbb{R}$. A function $\psi: \mathbb{R} \rightarrow \mathbb{C}$ is {\em square-integrable} if $\int_\mathbb{R}|\psi|^2d\mu < \infty$. Given two square-integrable functions $\psi: \mathbb{R} \rightarrow \mathbb{C}$ and $\varphi: \mathbb{R} \rightarrow \mathbb{C}$, we write $\psi \sim \varphi$ if $\int_\mathbb{R}|\psi-\varphi|^2d\mu=0$. Let $[\psi]$ be the set $\{\varphi: \varphi \sim \psi\}$. Let $L^2(\mathbb{R})$ be the Hilbert space of equivalence classes (mod $\sim$) of square integrable functions from $\mathbb{R}$ to $\mathbb{C}$, with inner product defined by $\langle [\varphi],[\psi] \rangle= \int_\mathbb{R}\overline{\varphi}\psi d\mu$.  (In what follows we will drop the square brackets, and understand all operations to be defined almost everywhere with respect to the Lebesgue measure on $\mathbb{R}$.  In other words, we will refer to an element $[\varphi]$ of $\mathscr{H}$ by a representative $\varphi$ of the equivalence class.)

There exists a representation of $U_a$ and $V_b$ on $\mathscr{H}=L^2(\mathbb{R})$.  We can express this representation explicitly as follows.  Define the operator $U_a$ to act on any vector $\varphi$ in $\mathscr{H}$ as $U_a\varphi(x)=e^{iax}\varphi(x)$, for all $x\in\mathbb{R}$; and define $V_b$ as $V_b\varphi(x)=\varphi(x+b)$, for all $x\in\mathbb{R}$.  These operators are bounded and defined on all of $\mathscr{H}$.  Moreover, they manifestly satisfy the Weyl form of the CCRs, since for all $\varphi\in L^2(\mathbb{R})$,
\[U_aV_b\varphi(x)=U_a\varphi(x+b)=e^{iax}\varphi(x+b)=e^{-iab}e^{ia(x+b)}\varphi(x+b)=e^{-iab}V_bU_a\varphi(x)\]  

The representation just defined is known as the \emph{Schr\"odinger representation} of the Weyl CCRs.  It has the following nice property: the maps $a \mapsto \langle \varphi , U_a \psi \rangle$ and $b \mapsto \langle \varphi , V_b \psi \rangle$ (which are just complex functions of one variable) are continuous for all $\varphi,\psi \in \mathscr{H}$.  Whenever a representation has this property, it is called \emph{regular}.  The Schr\"{o}dinger representation is the (essentially) unique representation of the Weyl CCRs with this property, in the following sense.  Two representations $( \{a\mapsto U_a \} , \{b\mapsto V_b \} )$ on $\mathscr{H}$ and $(\{a\mapsto U'_a \} , \{b\mapsto V'_b \} )$ on $\mathscr{H}'$ are \emph{unitarily equivalent} if there is a unitary transformation $W: \mathscr{H} \rightarrow \mathscr{H}'$ (i.e. a $W$ such that $\langle W \varphi, W \psi \rangle_{\mathscr{H}'} = \langle  \varphi, \psi \rangle_{\mathscr{H}}$ for all $\varphi, \psi \in \mathscr{H}$) such that $W \circ U_a \circ W^{-1} = U'_a$ and $W \circ V_b \circ W^{-1} = V'_b$.  We then have the following result:\footnote{For more on the Stone-von Neumann Theorem, see \citet{Ma49,Ri72,Pe90,Su99,ClHa01,Ruetsche}.}

\begin{thm}[Stone-von Neumann]If $(\{ U_a \} , \{V_b \} )$ is an irreducible\footnote{Recall that a representation $(\{U_a\},\{V_b\})$ is \emph{irreducible} just in case there are no nontrivial subspaces left invariant by all operators $U_a$ and $V_b$.} regular representation of the Weyl relations, then it is unitarily equivalent to the Schr\"odinger representation.\end{thm}

The Schr\"odinger representation has the following feature: there exist, on $\mathscr{H}$, \emph{unbounded} operators $Q$ and $P$, defined on a common dense subset of $\mathscr{H}$, that satisfy the CCRs in Eq. \eqref{CCR}; moreover, these operators take their standard form: for any vector $\varphi$ in their shared domains, $Q\varphi(x)=x\varphi(x)$ and $P\varphi(x)=-i\frac{\partial \varphi}{\partial x}$.  That these exist is a consequence of a basic result known as Stone's Theorem, which states that, given any (strongly) continuous one-parameter group of unitary operators $c\mapsto T_c$ on a Hilbert space, there exists a self-adjoint operator $O$ that ``generates'' the group in the sense that $T_c=e^{icO}$.  The Weyl operators $U_a$ and $V_b$ on $L^2(\mathbb{R})$ as we have defined them are strongly continuous one-parameter groups of unitary operators; indeed, the strong continuity of these one-parameter families is equivalent to the regularity condition.\footnote{Recall that a one parameter family of operators $c\mapsto T_c$ on $\mathscr{H}$ is \emph{strongly continuous} if, for every $\varphi\in\mathscr{H}$ and $c_0\in \mathbb{R}$, $\lim_{c\rightarrow c_0} ||T_c\varphi - T_{c_0}\varphi|| = 0$.  It is \emph{weakly continuous} if $c \mapsto \langle \varphi , T_c \psi \rangle$ is continuous for all $\varphi,\psi \in \mathscr{H}$.  Weak and strong continuity are equivalent for one parameter families of unitaries. Observe that regularity asserts that the one parameter families $a\mapsto U_a$ and $b\mapsto V_b$ are both weakly continuous, and thus also strongly continuous.}  Thus the generators $Q$ and $P$, respectively, of $a\mapsto U_a$ and $b\mapsto V_b$ exist, and are guaranteed to satisfy Eq. \eqref{CCR}.  

The operators $Q$ and $P$ in the Schr\"odinger representation can thus be taken to correspond (respectively) to the position and momentum magnitudes of a particle with one degree of freedom.  Moreover, it follows immediately from the Stone-von Neumann theorem and Stone's theorem that any \emph{other} regular representation of the Weyl CCRs will likewise admit operators $Q$ and $P$ satisfying Eq. \eqref{CCR}; and that, since any representation of $Q$ and $P$ will give rise, via exponentiation, to a strongly continuous representation of the Weyl CCRs, the Schr\"odinger representation is the unique representation, up to unitary equivalence, of the canonical commutation relations for $Q$ and $P$ as well.

We conclude this section by capturing the sense in which, in the Schr\"odinger representation, particles cannot have definite values for position or momentum.  By the spectral theorem \citep{ReSi80,KaRi97}, $Q$ is associated with a unique projection valued measure $E^Q$ on $\mathbb{R}$ such that
\[Q = \int_{\mathbb{R}}x\ dE^Q(x)\]
For any (Borel) measurable subset $S$ of $\mathbb{R}$, the projection $E^Q(S)$ is standardly interpreted as the projection associated with the proposition that the particle is in the region $S$.  As one can check, the projection $E^Q(S)$ can be defined on $L^2(\mathbb{R})$ by $E^Q(S)\varphi(x)=\chi_S(x) \varphi(x)$ where $\chi_S$ is the characteristic function of $S$. It follows that for all $\lambda \in \mathbb{R}$, $E^Q(\{\lambda\})=\textbf{0}$ (since for any $\varphi\in L^2(\mathbb{R})$, $E^Q(\{\lambda\})\varphi$ vanishes almost everywhere and hence $E^Q(\{\lambda\})\varphi\sim 0$).  Under the standard interpretation of projections, it follows that the truth value assigned to any proposition of the form ``the particle is located at the point $\lambda$'' is always ``false''. One standardly interprets this as the statement that a particle cannot be located at the point $\lambda$, which captures the claim with which we began the paper.  Similar claims hold for momenta.

\section{Halvorson on Non-regular Representations}\label{H}

In the previous section, we described the standard argument that takes one from the CCR, given by Eq. \eqref{CCR}, to the Schr\"odinger representation.  As we described, the Schr\"odinger representation is the unique regular representation, up to unitary equivalence, of the Weyl form of the CCRs; and in this representation, there is a precise sense in which particles cannot have definite values for position or momentum.  But as Halvorson points out, these claims rely on the regularity condition---and ``the regularity assumption begs the question against position-momentum complementarity'' \citep[p. 49]{HalvorsonBohr}.  In particular, if one considers representations of the Weyl form of the CCRs that fail to satisfy the regularity condition, the results described above fail. In fact, one can find representations in which particles \emph{can} have definite values for position or momentum \citep{beaume1974translation}.  

Let $\ell^2(\mathbb{R})$ denote the (non-separable) Hilbert space of square summable (as opposed to square integrable) functions from $\mathbb{R}$ into $\mathbb{C}$, defined as follows.  An element $\varphi$ of $\ell^2(\mathbb{R})$ is a complex-valued function $\varphi$ of a real variable that satisfies $\|\varphi\|=\sum_{x\in\Reals}|\varphi(x)|^2 < \infty$ (this implies $\varphi$ is supported only on a countable subset of $\mathbb{R}$).  The inner product on $\ell^2(\mathbb{R})$ is given by $\langle \varphi,\psi \rangle= \sum_{x\in\Reals}\overline{\varphi(x)}\psi(x)$.  For each $\lambda \in \mathbb{R}$, let $\chi_{\lambda}$ be shorthand for $\chi_{\{\lambda\}}$, the characteristic function of the singleton set $\{\lambda\}$. Then $\{\chi_\lambda:\lambda \in \mathbb{R}\}$ is an orthonormal basis for $\ell^2(\mathbb{R})$.

There exists a non-regular ``position'' representation of the Weyl form of the CCRs on $\ell^2(\mathbb{R})$. For all $a, b \in \mathbb{R}$, we introduce unitary operators $U_a$ and $V_b$ on $\ell^2(\mathbb{R})$ by defining their action on the orthonormal basis $\{\chi_\lambda:\lambda \in \mathbb{R}\}$ and extending linearly and continuously to all of $\ell^2(\mathbb{R})$: for each $\lambda \in \mathbb{R}$, let $U_a\chi_\lambda=e^{ia\lambda}\chi_\lambda$; and let $V_b\chi_\lambda=\chi_{\lambda-b}$.  One can verify that, for all $a,b \in \mathbb{R}$, $U_a$ and $V_b$, so defined, satisfy Eq. \eqref{WeylForm}.  

One can also check that, for all $\lambda\in\mathbb{R}$, $\norm{U_a \chi_\lambda -\chi_\lambda} \rightarrow 0$ as $a\rightarrow 0$, which implies $a \mapsto U_a$ is (strongly) continuous and, by Stone's theorem, there is an (unbounded) self-adjoint operator $Q$ on $\ell^2(\mathbb{R})$ such that $U_a=e^{iaQ}$ for all $a\in \mathbb{R}$.  Moreover, this operator $Q$ is such that for all $\lambda \in \mathbb{R}$, $Q\chi_\lambda=\lambda\chi_\lambda$.  In other words, the basis vectors $\chi_\lambda$ are eigenvectors for the position operator; we may likewise define a family of one dimensional spectral projections $E^Q_{\{\lambda\}}$ for $Q$.  Unlike in the Schr\"odinger representation, these projection operators do not vanish, and hence the proposition ``the particle has position $\lambda$,'' for some given $\lambda$, is not necessarily false.  In this sense, the position representation allows one to characterize states in which a particle has a definite position.  

But this representation has the following striking property: we cannot make sense of a momentum operator at all.  In particular, for any $\lambda\in\mathbb{R}$, $V_b \chi_\lambda$ is a unit vector orthogonal to $\chi_\lambda$ for all $b\neq 0$.  This means $\norm{V_b\chi_\lambda-\chi_\lambda} = 2$ for all $b\neq 0$, and hence $V_b\chi_\lambda$ does not converge to $\chi_\lambda$ as $b\rightarrow 0$.  Thus, $b \mapsto V_b$ is not strongly continuous, which implies that the representation is not regular.  It also means we cannot invoke Stone's theorem to find a generator of the unitary group $V_b$.  Indeed, as \citet{HalvorsonBohr} shows, there is no self-adjoint operator $P$ on $\ell^2(\mathbb{R})$ such that $V_b=e^{ibP}$ for all $b\in \mathbb{R}$.  Thus the momentum operator $P$ does not exist in this representation.

The same construction may be used, by switching the roles placed by $U_a$ and $V_b$, to produce a non-regular ``momentum'' representation.  In such a representation, there is a self-adjoint, unbounded operator $P$ on $\ell^2(\mathbb{R})$ that generates the one parameter group $V_b$.  This operator is a momentum operator; the vectors $\chi_\lambda$ are now understood as definite momentum eigenstates. But in this representation, for the same reasons as those just given, there is no self-adjoint operator $Q$ that generates $U_a$. In other words, the position operator $Q$ does not exist in this representation. 

Thus, if we relax the requirement of regularity, we can construct representations in which there are definite position states, and we can construct representations in which there are definite momentum states.  But the representation with definite position states does not have a momentum operator, much less definite momentum states; and the representation with definite momentum states does not have a position operator. In fact, these features of the representations we have just constructed turn out to be general.  We have:
\begin{thm}{\citep[Thm. 1]{HalvorsonBohr}} In any representation of the Weyl form of the CCRs, if $Q$ exists and has an eigenvector, then $P$ does not exist. If $P$ exists and has an eigenvector, then $Q$ does not exist.
\end{thm}
\noindent This is Halvorson's central result concerning (non-) regularity.

\section{The Physics of Non-regular Representations}\label{MM}

Halvorson draws attention to the non-regular representations we have discussed, and he indicates that they have some bearing on the interpretation of quantum theory.  He writes, for instance, that ``Indeed, by employing non-regular representations of the CCRs, we can make sense of Bohr's claims about position-momentum complementarity'' \citep[p. 55]{HalvorsonBohr}.  But he does not elaborate on this claim, and it is not clear how we are to understand the non-regular representations he discusses in a way that leads to a satisfying interpretation.  Here we will describe a few views one might adopt, and offer some considerations against accepting them. 

At first, when presented with multiple, unitarily inequivalent representations of the physical quantities of some quantum system,  one might look to the now-large literature on unitary inequivalence.  \citet{Ruetsche}, for instance, characterizes two dominant views in this literature: those of the \emph{Hilbert space conservative} and the \emph{algebraic imperialist}.\footnote{Ruetsche takes these terms from \citet{Arageorgis}; see also \citet{Fe17c}.  A third position is that of the so-called ``universalist'' \citep{Kronz+Lupher}, which may be thought of as a kind of middle ground between the algebraic imperialist and Hilbert space conservative positions, although \citet{Fe17c} argues universalism is equivalent to imperialism.}  The algebraic imperialist holds that one need not invoke Hilbert space representations of physical systems at all (except, perhaps, for convenience): the physical significance of a quantum theory is fully characterized by the abstract algebraic relationships between the magnitudes associated with that system, which are shared by all of the possible representations.  (In this case, that algebra would be the so-called \emph{Weyl algebra}, generated by $a\mapsto U_a$ and $b\mapsto V_b$; we return to this algebra below.)  Since Halvorson emphasizes the distinctive features of the position, momentum, and Schr\"odinger representations, however, it seems that the representations themselves are supposed to play some substantive role on his account.  And so, we set the algebraic imperialist position aside in this section.

The Hilbert space conservative, unlike the algebraic imperialist, holds that to interpret a quantum theory, one must first choose a Hilbert space and a collection of operators on that Hilbert space that represent the physically salient quantities.  The traditional view on which the Schr\"odinger representation offers a complete description of the possible states and magnitudes of the single free particle system is naturally construed as an instance of Hilbert space conservativism. The conservative's focus on Hilbert space representations suggests that this approach is closer to what Halvorson must have in mind.  But the Hilbert space conservative insists that there is a \emph{single} physically salient Hilbert space representation of a given physical system, such that the unit vectors in that Hilbert space represent the physically possible pure states of the system, and the self-adjoint elements of (some *-algebra of) the operators on that Hilbert space represent the observable quantities associated with that system.  It is not clear what to make of a proposal on which one has \emph{two} Hilbert space representations that are physically salient (or perhaps three---if we consider the Sch\"odinger representation).

So Halvorson's proposal does not fit naturally into the established frameworks for interpretation.  It seems that he wants to say that the physical significance of the one particle quantum system lies in its representations, but that one must consider multiple representations to fully describe the system.  We suggest two possible readings of this view.  

On one reading, to describe the full range of physical possibilities, we need to consider \emph{both} the position and momentum representations (and, perhaps, the Schr\"odinger representation), in the sense that the possible (pure) states of a single free particle correspond to the unit vectors in \emph{any} of these representations.\footnote{\label{fn:HSC}This can be understood as a Hilbert space conservative approach, wherein one uses the Hilbert space $L^2(\Reals)\oplus \ell^2(\Reals)\oplus \ell^2(\Reals)$, with the associated reducible representation of the Weyl form of the CCRs given by the direct sum of the Schr\"odinger, position, and momentum representations.  Hilbert space conservatives usually focus on irreducible representations, however, and so this proposal is unconventional.}  And the physical quantities correspond, say, to those that can be suitably constructed from the operators $U_a$ and $V_b$ acting on each of these Hilbert spaces.  For some, but not all, of these physical states---namely, those that happen to live in the position representation---one can define a definite position; for others, one can define a definite momentum.  This would capture a sense in which it is possible for a given physical system to possess a definite value for position or for momentum, but, if the system \emph{does} possess such a value for one of these quantities, no meaningful assertions can be made about the other quantity.

If this is the attitude we are meant to adopt, a number of questions arise.  First, one might expect that for systems in states  that happen to lie in one or the other representation, one can reason about those states in the ways we are accustomed to in non-relativistic quantum theories---as in, say, the Schr\"odinger representation.  For instance, one might expect that the Schr\"odinger equation will hold, for some Hamiltonian $H$ representing the free particle's energy.  This would imply that one could implement the dynamics for the free particle as a strongly continuous one parameter family $c\in\mathbb{R}\mapsto T_c=e^{icH}$ of unitary operators acting on those states.  But in fact, this cannot be done in all the representations under consideration.  

In particular, we will now show that there is a precise sense in which no suitably continuous free unitary dynamics can be defined on any representation with definite position states---including the position representation.  It is a direct consequence of this result that there is no Hamiltonian one can use to characterize the dynamics of the free one particle system in the position representation.  It then follows that there cannot be some Hamiltonian operator that describes, once and for all, the ``dynamics of the system'' independent of which representation the state of the system happens to fall in at a given time.\footnote{To re-express this claim in the terms of footnote \ref{fn:HSC}: there is no suitable unitary dynamics on the (reducible) representation described in that note.}

Let $M$ be a temporally oriented Galilean spacetime, i.e., an affine space of dimension $\geq 2$ endowed with temporal and spatial (degenerate) metrics $t$ and $h$, respectively \citep{MalamentGR,WeatherallSTG}.  (We have been considering one spatial dimension thus far, but we state this result more generally.)  In what follows, we say that vectors $\xi$ are \emph{timelike} if $t(\xi)\neq 0$; otherwise they are spacelike.   Now, consider structures of the form $(\mathscr{H}, p \mapsto E_p, \xi \mapsto T_\xi)$. Here, $\mathscr{H}$ is a Hilbert space. The map $p \mapsto E_p$ takes points in $M$ to projection operators on $\mathscr{H}$; the operator $E_p$ is supposed to represent the proposition that the particle is located at a particular point $p$ in space and time. The map $\xi \mapsto T_\xi$ is a unitary representation (not necessarily continuous) of the translation group on $M$. 

One would expect the position representation to give rise to a structure $(\mathscr{H}, p \mapsto E_p, \xi \mapsto T_\xi)$, as follows.  First, take $\mathscr{H}$ to be the Hilbert space of the representation.  This is a Hilbert space of definite position states in space,\footnote{One could equally well begin by considering a position representation on $l^2(\mathbb{R}^n)$, rather than $l^2(\mathbb{R}^{n-1})$, and consider states on spacetime.  The result below bears on this case, too.} and so some work will be necessary to associate it with points of spacetime.  Choose a point $o$.  Take $S$---space---to be the three dimensional affine space of points simultaneous with $o$, i.e., the points $o+\sigma$, for all spacelike vectors $\sigma$. Define, for $p\in S$, $p\mapsto E_p$ to take points $p$ to projections $E_p$ as defined in section \ref{H}. Finally, define spacelike translation operators through their action on definite position states associated with points in $S$, which, recall, form a basis for $\mathscr{H}$: take, for any spacelike vector $\sigma$, $T_{\sigma}$ to be the operator whose action on a definite position state $\chi_{p}$ is given by $T_{\sigma}\chi_p = \chi_{p+\sigma}$.  This operator simply takes the state representing a particle at point $p$ and returns the state representing a particle at point $p+\sigma$.  

Finally, we would like to extend this construction to spacetime.  If there were a suitable free dynamics in this representation, one would expect it to give rise to assignments $\xi\mapsto T_{\xi}$ of unitary operators corresponding to temporal evolution (time translation) in the timelike direction $\xi$.  For example, given a unit timelike vector $\xi$, one might na\"ively write down a Newtonian dynamics in the position representation by defining $T_{c\xi}\varphi(x) = \varphi(x-c\sigma)$ for $c\in\mathbb{R}$ the time elapsed. Here $\sigma$ is a spacelike vector representing the 3-velocity of the system relative to an observer with 4-velocity in the direction of $\xi$ (i.e., given a 4-velocity $\nu$ for the particle, $\sigma = \xi-\nu$).\footnote{We remark that this is a na\"ive proposal because it seems there is no way to associate a 4-velocity $\nu$ with such a particle.  What we are describing is a sketch of how one might \emph{try} to proceed; what we capture below is a sense in which there is no way to succeed.}  We would then define projection operators corresponding to points timelike related to $S$, by taking, for any $q\in M$, $q\mapsto E_q = T_{\xi}E_pT_{-\xi}$, where $p\in S$ and $\xi$ is such that $q = p + \xi$. And similarly we would define states by $\chi_q = T_\xi\chi_p$.  We would find, for instance, that for any vector $\eta$ and any point $q\in M$, we have $E_q\chi_q = \chi_q$ and $T_{\eta}\chi_q = \chi_{q+\eta}$.

We can motivate this sort of construction also by considering the situation in the Schr\"odinger representation. There, we can fix a unit timelike vector $\eta$ and define $T_{c\eta}=e^{-iP^2c/2}$, where $P$ is defined relative to $\eta$.\footnote{By this we mean that a vanishing expectation value for $P$ means that the expected velocity of the particle relative to an observer with 4-velocity $\eta$ is zero.  Alternatively, we may think of this as defining a Hamiltonian $H$ for which the average (kinetic) energy is determined relative to $\eta$.}  Acting with this operator ``translates'' the system in time by evolving its state, where the character of that evolution depends on the particle's momentum.

If we are given a single one-parameter unitary family $T_{c\eta}$ (where $\eta$ is a unit timelike vector and $c\in\mathbb{R}$) representing timelike translation along one direction, we can use it to reconstruct timelike translations along all other directions.  Given any other unit timelike vector $\xi$, we may define $T_{\xi} = T_{\eta}T_{\xi-\eta}$, where $\xi-\eta$ is necessarily spacelike; in other words, we translate the system in space, and then evolve.  Now, since we already have a suitable unitary operator $T_\sigma$ for each spacelike vector $\sigma$, we can make the choice $\sigma = \xi-\eta$ to use in the above definition of $T_\xi$ from $T_\eta$.  Observe that since we have required $\xi\mapsto T_{\xi}$ to be a representation of the entire translation group in spacelike and timelike directions (which is commutative), we must also have $T_{\xi}=T_{\xi-\eta} T_{\eta}$, i.e., translating in space and then evolving in time yields the same result as evolving in time and then translating in space.  This requirement makes sense only for dynamics that are not position dependent, e.g., free dynamics.\footnote{Observe that we consider only translations, not Galilean boosts.  The Hamiltonian itself does not change.}

Now consider the following three conditions. 
\begin{enumerate}
\item \textbf{Spatial Translation Covariance.} For all spacelike vectors $\sigma$ and all $p \in M$, $E_{p+\sigma}=T_\sigma E_pT_{-\sigma}$ (where $p+\sigma$ is the point that results from translating $p$ by the vector $\sigma$).
\item	\textbf{Exact Localizability.} If $p, q \in M$ are distinct and have vanishing temporal distance,\footnote{That is, $t(v)=0$, where $q=p+v$.} then $E_pE_q=\textbf{0}$.
\item \textbf{Timelike Continuity.} For all future-directed unit timelike vectors $\xi$, the (restricted one-dimensional) unitary representation $c\in\mathbb{R} \mapsto T_{c\xi}$ is weakly continuous.
\end{enumerate}

The first condition captures the idea that, whatever else is the case about the assignments $p\mapsto E_p$ and $\sigma\mapsto T_{\sigma}$, they satisfy a certain basic compatibility requirement under spatial translations.  We take this condition to be a weak necessary condition for the desired interpretation of the projections $p\mapsto E_p$ and operators $\sigma\mapsto T_{\sigma}$ under consideration.  It is trivially satisfied by the projection operators $p\mapsto E_p$ defined above for the position representation, since for any definite position state $\chi_q$, $E_{p+\sigma}\chi_q=\chi_q$ if $q=p+\sigma$ and zero otherwise; while $E_pT_{-\sigma} \chi_q = \chi_{q-\sigma}$ if $p=q-\sigma$ and zero otherwise, which means $T_\sigma E_pT_{-\sigma}=E_{q}$ if $q=p+\sigma$ (or, $p=q-\sigma$).  Observe that, though we are working in Galilean spacetime, we are not requiring covariance under the (full) Galilean group, just spatial translations.

The second condition captures the idea that the position states we are considering are ``exact'', in the sense that if a particle is located at a point $p$ at a time, then it cannot also be located at any (simultaneous) point $q$.  Note that this condition does indeed hold for the position projections defined in the position representation above.  Finally, the third condition captures the idea that the one-parameter unitary group determined by each timelike vector $\xi$ is suitably continuous.  This assumption, recall, is a necessary condition for the recovery of some unitary operator $H$ such that $T_{c\xi} = e^{icH}$ via Stone's theorem.  Note that we require continuity for all future-directed unit timelike vectors.  This is because we are working in Galilean spacetime, and so we presume that any suitable unitary dynamics could be expressed equally well for time translation along any temporal direction.  Going back to the example of free dynamics in the Schr\"odinger representation, described above, one should expect to be able to choose the vector $\eta$ relative to which time translation is defined arbitrarily---and, indeed, to define an operator $H$ relative to any standard of rest one likes.  

We can now formulate a proposition due to Malament (private communication) that shows a sense in which these conditions are incompatible.  (See \ref{app:a} for proof.)
\begin{prop}[Malament]\label{Malament} If a structure $(\mathscr{H}, p \mapsto E_p, \xi \mapsto T_\xi)$ satisfies conditions (1), (2), and (3), then $E_p=\textbf{0}$ for all points $p \in M$. \end{prop}
The conclusion of this proposition is naturally understood as the assertion that all propositions of the form ``the particle is located at $p$'' are (necessarily) false.  Since this condition is violated by any representation (including the position representation) that has definite position states, it follows that the position representation---and any other representation with definite position states---must violate one of the other assumptions.  Since the position representation $p\mapsto E_p$ does satisfy exact localizability, we conclude that there is no choice of maps $\xi\mapsto T_{\xi}$ in the position representation that satisfy both translation covariance and timelike continuity.  Moreover, we take translation covariance to be a necessary condition for the intended interpretation of the structure $(\mathscr{H},p\mapsto E_p, \xi\mapsto T_{\xi})$, and indeed translation covariance is satisfied for the spatial translations we defined in the natural way above for the position representation.  Thus, we conclude that there can be no satisfactory way to make time translation (or dynamics) weakly continuous in the position representation.

Halvorson himself says very little about dynamics in his paper, and he certainly never claims that the dynamics in the position (or momentum) representations arises from a Hamiltonian, or even that it is unitarily implementable.  What he \emph{does} say, in a footnote on the penultimate page of the article, is that the standard (free) Schr\"odinger dynamics ``can also be defined in a representation-independent manner, at least for the standard case of a free particle'' \citep[pg.~55]{HalvorsonBohr}.  Halvorson expresses this dynamics directly via its action on the Weyl operators $U_a$ and $V_b$, as:\footnote{See also \citep{FaVe74,MR1213612}.  Observe that this transformation is not precisely what Halvorson himself writes.  Halvorson's expression does not involve the phase factors we include here, and as a result his transformation does not correspond to the standard free particle dynamics in the Schr\"odinger representation.}
\[
U_a V_b\mapsto e^{\frac{ica^2}{2}}U_a V_{b+ca}.
\]
One can confirm that this transformation is unitarily implementable in the Schr\"odinger representation, in the sense that it determines a (strongly continuous) one-parameter family of unitary operators $c\mapsto T_c$, and moreover, that the generator of the group is the standard free Hamiltonian $H=P^2/2$.  

This proposal does seem like the natural way to generalize the standard Schr\"odinger dynamics to the position and momentum representations.  But it has some unattractive consequences.  In particular, although this dynamics is also unitarily implementable in the momentum representation, in precisely the same sense as in the Schr\"odinger representation, it is \emph{not} unitarily implementable in the position representation \citep[Prop. 3.27]{beaume1974translation}.\footnote{Observe the relationship between this claim and that in Prop. \ref{Malament}.  There we show that no unitarily-implementable dynamics satisfying certain basic desiderata is continuous; now we are claiming that a particular proposal for a free dynamics is not even unitarily implementable, irrespective of continuity properties.}  Thus we cannot understand Halvorson's proposed dynamics as an instance of unitary dynamics at all, at least for position states.

Worse, recall that we are considering the idea that the states in both the position and momentum representations (and, perhaps, the Sch\"odinger representation) are physically possible states, so that a given particle may be in a definite position state or it may be in a definite momentum state.  One might have imagined that this setup would allow us to consider particles that are sometimes in definite position states and sometimes in definite momentum states, in the sense of being able to evolve from one such state to another.  But the facts just noted about (non-)unitary dynamics show that this is not possible, at least for the proposed free dynamics.  Instead, if at any particular time, a particle is in a definite momentum state, and it then evolves according to the free dynamics (or any other dynamics that is unitarily implementable in the momentum representation), then at all \emph{other} times, the particle must \emph{also} be in a state in the momentum representation.  Likewise for states in the Schr\"odinger representation.  This is because any dynamics that is unitarily implementable in a given representation takes states in that representation only to other states in that same representation.  

This consequence does not hold for the position representation, since there the dynamics is not unitarily implementable.  But in a sense, the situation there is even stranger.  In fact, states in the position representation never remain position states: they \emph{necessarily} evolve, in any finite time under the free dynamics, to states in other representations \citep[Prop. 3.27]{beaume1974translation}.  But they cannot evolve to any state in the momentum or Schr\"odinger representations! In other words, the states in the various representations under consideration are not mutually dynamically accessible, at least under the free dynamics; and states in the position representation immediately evolve to states that do not lie in any of the three representations we have discussed.  This conclusion is in tension with the idea that it is the states in precisely these representations that one should take to have physical significance.

Of course, these remarks hold only for free dynamics.  And in that context, the mutual dynamic inaccessibility of position and momentum states may not be so surprising or troubling---after all, a free particle with definite momentum, say, might be expected to propagate without change to its momentum, and thus never evolve into a definite position state.  In other words, one might have thought that some sort of non-trivial interaction would be necessary to have a momentum state evolve into a position state and vice versa.  

Fair enough. But the underlying facts that lead to the features of the free dynamics just discussed can be expected to cause problems more generally.  First, consider that \emph{any} dynamics that is unitarily implementable in the Schr\"odinger representation, including dynamics encoding interactions and potentials, will be such that it takes states in the Schr\"odinger representation only to states in the Schr\"odinger representation.  This means that none of the interactions one ordinarily considers in quantum theory could evolve a particle that happens to be in a Schr\"odinger state at some time into a position or a momentum state at other times.  On the other hand, general Hamiltonians, which in the Schr\"odinger representation would be represented as polynomials of both $P$ and $Q$, generally cannot be expressed in the position and momentum representations by one-parameter families of unitary operators.  Indeed, such dynamics generally cannot even be understood as automorphism groups acting on the Weyl algebra \citep{FaVe74}, so retreating to an algebraic perspective, as we indicate above Halvorson seems to suggest, does not help  here.

All of this discussion of dynamics has been premised on the idea that we should think of the position and momentum representations in something like the way that a Hilbert space conservative would usually proceed, taking all and only the states in these representations to be physically possible.  We have now seen some reasons to think this approach is unattractive.  But we mentioned above that another interpretation of Halvoron's view is available.  This interpretation is suggested by Halvorson's invocation of Bohr's philosophy---and by his remark that ``one representation cannot be preferable to another on empirical grounds alone'' \citep[p. 55]{HalvorsonBohr}, which seems to imply that some other, non-empirical consideration leads us to choose between representations.  In the context of the paper, then, one might take him to be suggesting that these extra-empirical considerations may vary with circumstances, context, or perspective, in such a way that which representation one should use to describe a given quantum system is contextual or perspectival. 

Our discussion of this possibility will be brief, because Halvorson does not elaborate it (or clearly endorse it).  But we take the basic idea to be that, contra the Hilbert space conservative, we should not suppose that the position and momentum representations precisely delimit the possible physical states of a single particle in any literal sense; rather, they provide resources that we may use, in some contexts and for some purposes, to describe the actual physical states of affairs in some approximative way.  And so, given some objective physical situation, one might choose to describe it using a state in the position representation; or one might choose to describe it using a state in the momentum representation.  The considerations that guide this choice, on this view, are not fixed by the features of the system itself.  Most importantly, one should not suppose that there is any absolute fact of the matter about ``the'' vector representing a particle's state in \emph{any} of these representations.

There are some disadvantages to this approach, such as the fact that it seems to make mysterious what the underlying facts are regarding a system that support the choice of a given state to represent it.  In particular, it provides no insight into what the structure of the space of states for the particle actually is; it tells us only that different state spaces, with ambiguous relations to one another, can be used.  Perhaps worse given how Halvorson frames his paper, this attitude tells us nothing about whether quantum particles do or can possess definite positions or momenta---merely that they may be described as having such properties, from some perspective or in some contexts.  Finally, the concerns we have already raised about dynamics arise on this view as well, since it is not clear how we should understand changes from contexts in which one would use a position state (say) to ones in which one would use a momentum state.  Are these changes merely in the perspective of an external observer, or are there facts about a given physical system (say, its interactions with certain ``classical'' measuring instruments) that play a role?

We will not try to attribute answers to these questions to Halvorson.  Instead, we will move on to consider a different interpretation of Halvorson that lies in the same vicinity, but which is a bit more precise.  On this alternative proposal, the representations we have been considering are, in a sense, secondary; their value comes from the insight they give into the physical content of a certain algebra of quantities.  This algebra may be studied independently of any representations at all---and it can also provide a framework for better understanding the relationship between the representations we have been considering.

\section{Regularity Revisited}\label{F}

Given the difficulties of the previous section with providing an interpretation of non-regular representations that does not stray too far from the traditional Hilbert space conservative approach, we now consider if an algebraic imperialist approach can shed any light.\footnote{The algebraic imperialist approach considered here differs somewhat from that described by \citet{Ruetsche}, which is tied to what she calls ``pristine interpretation''.  For the purposes of this section, an algebraic imperialist approach is one that uses broadly algebraic tools in the abstract without focusing on particular representations, but is not necessarily pristine.}  In fact, such an algebraic approach might even come closer to Halvorson's view, expressed in the following remark:
\begin{quote}
The abstract Weyl algebra carries the full empirical content of the quantum theory of a single particle.  In particular, the Weyl algebra has enough observables to describe any physical measurement procedure and enough states to describe any laboratory preparation.  A representation does not make any further empirical predictions; indeed, it adds \emph{nothing} in the way of empirical content. \citep[][p. 55]{HalvorsonBohr}
\end{quote}
The purpose of this section is to investigate the plausibility of such an interpretation. 
We will present a reading of Halvorson, and then describe the outline of an argument that an algebraic approach does \emph{not} justify the use of non-regular representations---at least, not without substantial further assumptions that we feel are not warranted.  A proper treatment of this subject, however, requires much more space and a different set up than we feel is appropriate for the present paper.  And so, this section aims only to isolate the central issues, which we will then take up in a companion paper.

On an algebraic approach to quantum theory, rather than constructing a quantum theory by identifying a Hilbert space and a subalgebra of the (bounded) operators acting on that Hilbert space, one instead begins by considering just an abstract C*-algebra of quantities, and then proceeds to define other notions, such as states, using this algebra.\footnote{More precisely, a C*-algebra $\mathfrak{A}$ is an involutive, associative, complete normed algebra satisfying the C* identity, $||A^*A||=||A||^2$ for all $A\in\mathfrak{A}$; and the \emph{state space} of a C*-algebra $\mathfrak{A}$ is the colllection of all positive, normalized, complex-valued linear functionals $\omega:\mathfrak{A}\rightarrow\mathbb{C}$.} One can think of the approach as abstracting away from any particular ``representation'' of the system and trying to characterize the abstract relationships between physical quantities that all such representations have in common. Note that there is some abuse of language in our use of the term ``representation'' here, as a representation of a C*-algebra is a homomorphism of the algebra into the bounded operators on some Hilbert space that preserves the involution operation; whereas previously we have used the term ``representation'' in a more general sense that has not required us to define an algebra.  But for our purposes, the two senses are essentially interchangeable, and no ambiguities will arise.  

In the remarks above, Halvorson invokes a particular algebra of quantities: the abstract Weyl algebra $\mathcal{W}$ \citep{MaSiTeVe74,Pe90}.  The definition of the abstract Weyl algebra is not necessary for present purposes; we discuss it in more detail in the sequel \citep{Feintzeig+WeatherallPart2}.  What is important is that it is a C*-algebra that is determined, via closure under algebraic operations, from the Weyl unitaries $U_a$ and $V_b$ discussed above.  In this sense, it may be thought of as an algebra that has been in the background all along in our discussion of the Weyl form of the CCRs and in representations thereof.  Indeed, any representation of the Weyl CCRs, in the sense of the previous sections, yields a representation of the abstract Weyl algebra, and vice versa.  Thus everything that has been said previously about representations of the CCRs for $U_a$ and $V_b$---including what we have said about regular and non-regular representations---extends unchanged to representations of the Weyl algebra $\mathcal{W}$.  

What advantages might one find by moving from the particular representations discussed in the previous sections to the the abstract Weyl algebra?  Once we choose a representation, some, but not all, of the states on $\mathcal{W}$ will correspond to density operators in that representation; likewise, some, but not all, of the pure states will correspond to rays in the Hilbert space.  (The collection of states that have a density operator representative in a representation is call the \emph{folium} of the representation.)  In general, though, any density operator in any representation of an abstract C*-algebra $\mathfrak{A}$ determines a state on that algebra.  And thus, retreating to the abstract Weyl algebra allows one to consider all of the states in all three of the representations described above as on a par: they are all states on the same algebra.  This is presumably what Halvorson means when he says that the abstract Weyl algebra has all of the resources needed to describe any observables and any experimental preparation involving a single particle.\footnote{To be sure, the observables $P$ and $Q$ are \emph{not} to be found in the Weyl algebra; as we discuss in the previous section, one gets each of these only in some representations, but not others.  So it is perhaps question begging to assert that \emph{all} physically relevant observables are to be found in the Weyl algebra.}

Thinking of the states in these representations as states on the same algebra also allows one to characterize certain relationships between them.  In particular, a well known result known as Fell's theorem \citep{Fe60,Kronz+Lupher} establishes a certain precise sense in which the states of the Schr\"odinger, position, and momentum representations of the Weyl algebra all approximate one another arbitrarily well, in a particular topology determined by the abstract Weyl algebra.   

More precisely, one can use the abstract Weyl algebra to define a topology, known as the weak* topology, on its state space.  Fell's theorem establishes that the folium of any faithful representation of a C*-algebra is dense in the folium of any other faithful representation in the weak* topology on the state space.  In other words, every state with a density operator representative in one faithful representation can be approximated as a limit in the weak* topology of states with density operator representatives in another faithful representation.  Now it suffices to notice that the Schr\"odinger, position, and momentum representations are all faithful,\footnote{In fact, since the Weyl algebra is \emph{simple}, all of its nontrivial representations are faithful.} which establishes that the corresponding states in each folium can all be used to approximate one another.

The above sketch of an argument provides one sense in which the Schr\"odinger, position, and momentum representations of the Weyl algebra might be understood to be on a par, at least with respect to the collection of states they appear to deem physically possible.  This is at least one way of making precise Halvorson's claim that choosing a particular representation adds no empirical content: any one of these choices has the resources to represent, to whatever degree of approximation one likes, precisely the same situations as any other choice.  One can also see, here, a sense in which one might freely choose states from different representations, depending on context, along the lines of the second reading of Halvorson proposed in section \ref{Malament}.  In our view, this is the most compelling interpretation of Halvorson's proposal; even if it is not the interpretation Halvorson intended, we believe it is one worth considering. 

Does this mean that Halvorson's proposal is vindicated, and that non-regular representations have the same status as the Schr\"odinger representation?  We think this conclusion is too fast.  In particular, the interpretation of Halvorson that we have just given relies heavily on the use of a particular topology: the weak* topology on the state space of $\mathcal{W}$.  For the interpretation to go through, the weak* topology must provide a physically relevant notion of approximation for the empirical content of states.\footnote{Here we echo a point made forcefully by \citet{Fl16}, that different choices of topology encode different senses of approximation---and in any particular case, careful attention must be paid to whether a particular topology captures the salient sense.  See also \citet{Fe17c} for applications of Fletcher's ideas to quantum theories.}  But does it?  

It is here that we begin to just sketch an argument; the full details of this line of argument appear in the companion paper \citep{Feintzeig+WeatherallPart2}.  We remark first that the weak* topology on the state space of the Weyl algebra is determined by the Weyl algebra itself.  This means that whether the weak* topology is physically significant is deeply intertwined with whether the Weyl algebra itself has the physical significance that Halvorson attributes to it, as the natural or correct algebra of quantities for a one particle system.  We emphasize that the Weyl algebra is not the only possible choice---and more, mathematical physicists and philosophers have presented a number of arguments that it is the \emph{wrong} choice \citep[see, e.g.,][]{FaVe74,La90,La90b,BuGr08,GrNe09,Fe18}.  Moreover, if one were to make a different choice, one would arrive at a different state space, with a different weak* topology, and a different verdict on whether, for instance, there are definite position and momentum states that are dense in the folium of the Schr\"odinger representation (suitably understood).  Thus, at very least, the defender of Halvorson's view, on the interpretation we have proposed in this section, is obligated to justify the choice of the abstract Weyl algebra.

In fact, once one questions the choice of the Weyl algebra as a starting point, a new and deeper problem emerges for Halvorson's proposal: not all algebras that one might choose will allow for non-regular states or non-regular representations at all.  For example, \citet{La90b} provides a construction procedure for an algebra that does not admit any non-regular representations \citep[see also][]{GrNe09}.  In other words, the very existence of the sorts of definite position and momentum states that Halvorson urges us to consider depends on the initial choice to consider representations of the Weyl algebra---or rather, the choice to consider representations of the Weyl form of the CCRs.

Of course, for the purposes of engaging with Halvorson's proposal, it would be question-begging to simply choose some other algebra that did not allow for non-regular states or representations without some justification.  But the converse is also true: to start with the Weyl algebra without further argument---and then claim that, because it admits non-regular states, those states must have physical significance---is also to beg the question.  The principal issue, then, concerns which algebra to begin with.

Having isolated, then, what we take to be the main issue in evaluating Halvorson's proposal, we will defer a detailed discussion of how to settle that issue to the sequel paper.  We remark only that we will adopt there a particular attitude towards how to construct quantum algebras, previously defended by \citet{Fe17b}.  And we will prove a result with the following interpretation: any acceptable representation of a suitably chosen quantum algebra is necessarily regular.  Of course, a lot of work is done by the terms ``acceptable'' and ``suitably chosen'' here, and to make clear what we mean requires significantly more discussion.  But the upshot will be that the justifications for these choices do not run via a prior assumption of regularity, or a prior rejection of definite position or momentum states.  

\section{Conclusion}\label{conclusion}

The restriction to regular representations of the Weyl CCRs (or, respectively, the abstract Weyl algebra) for a single particle is standard in textbook treatments of quantum theory.  It is also widely taken for granted in the philosophical literature: consider, for instance, that any claim to the effect that quantum field theory faces a distinctive problem of ``unitarily inequivalent representations'' \citep{Arageorgis,Ruetsche} relies on the fact that the Stone-von Neumann theorem guarantees, for finite-dimensional systems,\footnote{Really, there is an extra caveat here that the Stone-von Neumann theorem only applies to systems with simply connected phase space $\mathbb{R}^{2n}$.  For systems with finitely many degrees of freedom (and hence, without going as far afield as field theories) that have topologically nontrivial phase spaces, unitarily inequivalent representations that are, in some sense, regular may still arise \citep{La90,La90b}.} an essentially unique representation of the Weyl algebra---a theorem that, we have seen, holds only in the presence of the regularity assumption.  Regularity is so standard, in fact, that it is rarely justified---or questioned.  It is treated as a harmless technical assumption, ruling out pathologies of no physical interest.  

Equally standard, of course, is the claim that a quantum particle cannot have a definite position or momentum.  This latter claim, however, has manifest physical significance, and it bears on our basic understanding of physical magnitudes in quantum theory.  And as \citet{HalvorsonBohr} correctly points out, the (non-)existence of definite position and momentum states turns precisely on the status of the regularity assumption.  If regularity is assumed, definite position and momentum states are not physically possible; conversely, if regularity is rejected, one can recover these states in the position and momentum representations.  Far from a benign technical assumption, regularity rules out possibilities that might have direct bearing on the interpretation of quantum theory.

It seems clear, then, that the status of regularity deserves far closer attention than it is usually given.  The present paper has taken up one part of the project of scrutinizing this condition.  We have focused here on Halvorson's arguments concerning what one gets if one rejects regularity---namely, definite position and momentum states that, he argues, may provide some theoretical grounding for Bohr's principle of complementarity.  

We have discussed three different readings of Halvorson's proposal.  Two of these readings follow Halvorson's own emphasis on the non-regular position and momentum representations as the main upshot of dropping regularity.  We argued, however, that neither of these options is appealing.  The principle difficulties have to do with dynamics.  On the one hand, we presented Malament's result, and argued that it shows there is no suitable free dynamics available in the position representation.  On the other hand, we studied the consequences of Halvorson's own proposal to retreat to dynamics that may be expressed in a representation-independent (algebraic) manner, and argued that those dynamics are not unitarily implementable in the position representation.  We went on to argue that in general, the three representations Halvorson considers are dynamically inaccessible, which creates serious interpretational problems if we wish to understand the states in all three of them as having physical significance.

Our third reading of Halvorson's proposal had a different flavor: it was more algebraic in character, invoking Fell's theorem to capture a sense in which the states in each of the position, momentum, and Schr\"odinger representations may be understood to approximate one another arbitrarily well.  This result underwrote a suggestion that we should be free to work with whichever representation is most convenient for some purpose---including the position and momentum representations---because whatever state a physical system in fact occupies can always be approximated as well as one likes by a position state, a momentum state, or a Schr\"odinger state.  This suggests there is nothing at stake in the choice between representations---a result that does seem evocative of Bohr's principle of complementarity.  One might take this proposal as a friendly elaboration of Halvorson's view; we are not sure it is precisely what he had in mind, but we think it is worthy of serious reflection.

But as we also argued, albeit briefly, whether this proposal succeeds depends on a prior choice, concerning which abstract algebra to use to represent a quantum particle.  And so to properly defend this view, one needs to justify the choice of the Weyl algebra for representing the physical quantities of the system considered.  But as we suggest---and elaborate in the sequel---there are good reasons to choose a \emph{different} algebra to represent the physical magnitudes associated with a quantum particle.  And alternative choices of algebra lead to different consequences concerning whether there are non-regular representations at all.  

Of course, we do not make this case in the present paper; we leave that to the sequel.  What we have done, however, is show that there is more to the story of the regularity assumption than Halvorson suggests.  It is not just that there is a close connection between regularity and the (im)possibility of definite position and momentum states; in fact, both of these are also deeply related to how one goes about constructing a quantum theory in the first place.  This moral may be understood in two ways: one is that it makes clear why there is a great deal of physics involved in judiciously choosing an algebra, since basic issues---such as whether definite positions and momenta are possible---turn on this choice; or, conversely, it provides a new way of understanding (and, perhaps, justifying) regularity as associated with the algebra we choose.  Pursuing these lines is the goal of the next paper in this series.

\appendix 

\section{Proofs of Propositions in \S\ref{MM}}
\label{app:a}

We will first need a lemma, which uses the following notion:\\

\noindent (4) \textbf{Spacelike Continuity.} There is a unit spacelike vector $\sigma$ such that the (restricted one-dimensional) unitary representation $c \mapsto U_{c\sigma}$ is weakly continuous (i.e. the function $c \mapsto \langle \psi, U_{c\sigma} \varphi \rangle$ is continuous for all states $\psi, \varphi \in \mathscr{H}$). \\

\noindent \textbf{Lemma.} If a structure $(\mathscr{H}, p \mapsto E_p, \xi \mapsto U_\xi)$ satisfies conditions (1), (2), and (4), then $E_p=\textbf{0}$ for all points $p \in M$. \\

\noindent {\em Proof.} Assume conditions (1), (2), and (4) hold. Let $\sigma$ be as in (4) and let $p$ be any point in $M$. For all $c\neq 0$, $p + c\sigma$ and $p$ are distinct and in the same simultaneity slice. By condition (2), $E_pE_{p+c\sigma}=\textbf{0}$. But $E_{p+c\sigma}=U_{c\sigma}E_pU_{-c\sigma}$ by condition (1). So $E_pU_{c\sigma}E_pU_{-c\sigma}=\textbf{0}$ and therefore $E_pU_{c\sigma}E_p=\textbf{0}$ for all $c\neq 0$. Now, for any $\psi, \varphi \in \mathscr{H}$, $\langle E_p \psi, U_{c\sigma}E_p \varphi \rangle = 0$ for all $c \neq 0$.  By condition (3), $\langle E_p \psi, U_{c\sigma}E_p \varphi \rangle = 0$ for $c= 0$ as well. So, $\langle E_p \psi, E_p \varphi \rangle=0$. And because $E_p$ is a projection operator, we have $\langle \psi, E_p \varphi \rangle = \langle \psi, E_pE_p \varphi \rangle = \langle E_p \psi, E_p \varphi \rangle=0$. Since $\psi$ and $\varphi$ are arbitrary, it follows that $E_p=\textbf{0}$. $\square$\\

\noindent {\em Proof of Prop. \ref{Malament}.} Assume $(\mathscr{H}, p \mapsto E_p, \xi \mapsto U_\xi)$ satisfies conditions (1), (2), and (3). Let $\xi$ and $\xi'$ be distinct future-directed unit timelike vectors. By condition (3), the (restricted one-dimensional) unitary representations $c \mapsto U_{c\xi}$ and $c \mapsto U_{c\xi'}$ are weakly continuous. Consider the vector $\sigma=\xi'-\xi$. It is spacelike and non-zero. If $\sigma$ satisfies condition (4), it follows that the normalized vector $\sigma/\|\sigma\|$ does as well. Therefore, we are done (by the lemma above) if we can show that $\sigma$ satisfies condition (4).

Consider the function $g(c)=\langle \psi, U_{c\sigma} \varphi \rangle$ for any $\psi, \varphi \in \mathscr{H}$. Clearly, if $g(c)$ is continuous at $c=0$, it is continuous everywhere. Let $\psi, \varphi \in \mathscr{H}$ be arbitrary. Because $\langle \psi, U_{c\sigma} \varphi \rangle=\langle \psi, U_{c\xi'} U_{-c\xi} \varphi \rangle=\langle U_{-c\xi'} \psi,  U_{-c\xi} \varphi \rangle$, we have the following.

\begin{eqnarray*}
|g(c)-g(0)|&=&|\langle U_{-c\xi'}\psi, U_{-c\xi}\varphi \rangle - \langle \psi, \varphi \rangle |\\
&=&|\langle U_{-c\xi'}\psi, U_{-c\xi}\varphi \rangle - \langle \psi, U_{-c\xi}\varphi \rangle + \langle \psi, U_{-c\xi}\varphi \rangle - \langle \psi, \varphi \rangle |\\
&\leq&|\langle U_{-c\xi'}\psi-\psi, U_{-c\xi}\varphi \rangle + \langle \psi, U_{-c\xi}\varphi-\varphi \rangle |\\
&\leq&\|U_{-c\xi'}\psi-\psi\|\|U_{-c\xi}\varphi\|+\|\psi\|\|U_{-c\xi}\varphi-\varphi\|
\end{eqnarray*}

Because $U_{-c\xi'}$ is unitary, we have the following.

\begin{eqnarray*}
\|U_{-c\xi'}\psi-\psi\|^2&=&\langle U_{-c\xi'}\psi, U_{-c\xi'}\psi \rangle - \langle U_{-c\xi'}\psi, \psi \rangle - \langle \psi, U_{-c\xi'}\psi \rangle + \langle \psi, \psi \rangle\\
&=&2\langle \psi, \psi \rangle- \langle \psi, U_{c\xi'}\psi \rangle - \langle \psi, U_{-c\xi'}\psi \rangle\\
\end{eqnarray*}

Since $c \mapsto U_{c\xi'}$ is weakly continuous, both $\langle \psi, U_{c\xi'}\psi \rangle$ and $\langle \psi, U_{-c\xi'}\psi \rangle$ converge to $\langle \psi, \psi \rangle$ as $c \rightarrow 0$. Thus, $\|U_{-c\xi'}\psi-\psi\|$ converges to zero as $c \rightarrow 0$. By a parallel argument, so does $\|U_{-c\xi}\varphi-\varphi\|$. Finally, because $U_{-c\xi}$ is unitary, $\|U_{-c\xi}\varphi\|=\|\varphi\|$. So, $|g(c)-g(0)|$ converges to zero as $c \rightarrow 0$. $\square$\\

\section*{Acknowledgments}
We are grateful to Hans Halvorson, David Malament, and Laura Ruetsche for helpful conversations related to this material.

\singlespacing

\bibliographystyle{elsarticle-harv}
\bibliography{regularity}
\end{document}